\documentclass[5p,twocolumn,times,number]{elsarticle}
\usepackage{graphicx}
\usepackage{subfigure}

\begin{document}

\title{Development of Muon Drift-Tube Detectors for High-Luminosity Upgrades of the Large Hadron Collider}

\author{B.~Bittner}\author{J.~Dubbert}\author{O.~Kortner}\author{H.~Kroha\corref{cor}} 
\ead{kroha@mppmu.mpg.de}
\author{F.~Legger}\author{R.~Richter}
\address{Max-Planck-Institut f\"ur Physik, F\"ohringer Ring 6, D-80805 Munich, Germany} 
\author{O.~Biebel}\author{A.~Engl}\author{R.~Hertenberger}\author{F.~Rauscher}
\address{Ludwig-Maximilians-Universit\"at M\"unchen, Am Coulombwall 1, D-85748 Garching, Germany}
\cortext[cor]{Corresponding author. Tel.: +49-89-32354-435, fax: +49-89-32354-305.}

\begin{abstract}
The muon detectors of the experiments at the Large Hadron Collider (LHC) 
have to cope with unprecedentedly high neutron and gamma ray background rates.
In the forward regions of the muon spectrometer of the ATLAS detector, for instance,
counting rates of 1.7~kHz/cm$^2$ are reached at the LHC design luminosity.
For high-luminosity upgrades of the LHC, up to 10 times higher
background rates are expected which require replacement of the muon chambers
in the critical detector regions. Tests at the CERN Gamma Irradiation Facility
showed that drift-tube detectors with 15 mm diameter aluminum tubes 
operated with Ar:CO2 (93:7) gas at 3 bar and a maximum drift time of about
200 ns provide efficient and high-resolution muon tracking up to the highest 
expected rates. For 15~mm tube diameter, space charge effects deteriorating
the spatial resolution at high rates are strongly suppressed.
The sense wires have to be positioned in the chamber with an accuracy
of better than $50~\mu$m in order to achieve the desired spatial resolution of a chamber 
of $50~\mu$m up to the highest rates.
We report about the design, construction and test of prototype detectors
which fulfill these requirements.
\end{abstract}

\begin{keyword}
Drift tubes \sep muon chambers \sep LHC
\end{keyword}

\maketitle

\section{Introduction}

The muon detectors of the experiments at the Large Hadron Collider (LHC) 
will encounter unprecedentedly high background counting rates due to neutrons and gamma rays
in the energy range up to about 10~MeV which originate mainly from secondary interactions
of the hadronic collision products with accelerator elements, shielding material and the
detector components. The forward regions of the detectors are particulary exposed to the
background radiation.

In the muon spectrometer of the ATLAS detector~\cite{ATLAStp,ATLASpaper} at the LHC,
large Monitored Drift-Tube (MDT) chambers are used for precision tracking 
in a toroidal magnetic field of superconducting air-core magnets~\cite{MuonTDR,ATLASpaper}.
The MDT chambers consist of two triple or quadruple layers of pressurized
aluminum drift tubes of 30~mm outer diameter and 0.4~mm wall thickness, filled with an Ar:CO$_2$ (93:7) gas mixture at an
absolute pressure of 3~bar. An operating voltage of 3080~V, corresponding to a gas gain of $2\cdot 10^4$, is applied 
between the tube wall and the 50~$\mu$m diameter anode wire. The gas mixture and gas gain have been chosen
to prevent aging of the drift tubes up to an accumulated charge of at least 0.6~C/cm in the high background
environment at the LHC.
The average spatial resolution of individual drift tubes at low background rates of 80~$\mu$m together with the 
positioning accuracy of the sense wires in a chamber of $20~\mu$m
translates into a spatial resolution of a MDT chamber of 35~$\mu$m. 

The highest background rate in the ATLAS MDT chambers at the LHC design luminosity of
10$^{34}$cm$^2$s$^{-1}$ is expected to be 100~Hz/cm$^2$ in the inner endcap layers closest to the
beam pipe~\cite{baranov}. The limited knowledge of the showering process in the absorber, chamber sensitivities and
cross section and particle multiplicity of the primary proton collisions at the LHC center-of-mass energy of $\sqrt{s} =
14$~TeV is taken into account in a safety factor of 5.
Hence the MDT chambers are designed to cope with particle fluxes of up to 500~Hz/cm$^2$ corresponding to a
maximum counting rate of 300~kHz in 2~m long drift tubes of the inner forward chambers.

The LHC upgrade schedule foresees a continuous luminosity increase up to three times the design luminosity 
and followed eventually by a larger upgrade to ten times the design luminosity called Super-LHC (S-LHC). 
Assuming that the background rates will scale with the luminosity, the degradation of the performance of the MDT
chambers will compromise the ATLAS physics goals. We investigate the possibility of using drift-tube detectors
with smaller tube diameter and therefore shorter drift-time for the regions of highest background rates
in the muon detectors of the LHC experiments. Building on the experience with the ATLAS MDT chambers,
new muon drift-tube detectors with 15~mm diameter tubes for counting rates up to about 1.5 kHz/cm$^2$ 
have been developed and tested.

\section{Drift-tube performance at high rates}

At high counting rates, the drift tubes of the MDT chambers are known to suffer 
from a degradation of the spatial resolution
due to space-charge effects~\cite{aleksa,deile} and of the muon detection efficiency 
due to the increased drift tube occupancy~\cite{sandra}.
Both effects can be supressed by reducing the tube diameter while leaving the other
parameters of the drift tubes, in particular the gas mixture and pressure and the gas gain, unchanged.

\begin{figure}[h]
\centering
\includegraphics[width=0.5\textwidth]{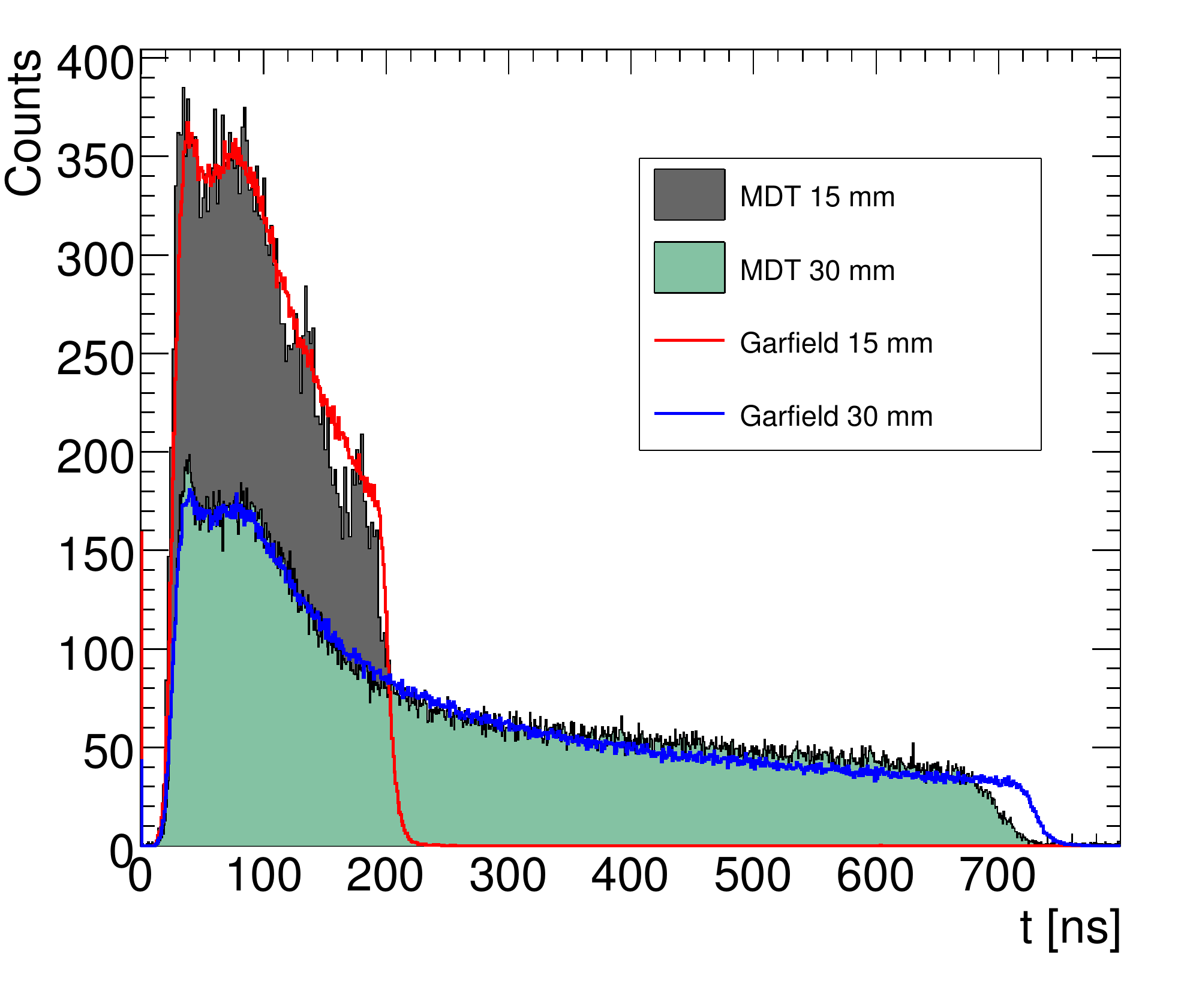}
\caption{Drift-time spectra of 30~mm and 15~mm diameter drift tubes operated with Ar:CO$_2$ (93:7) gas
mixture at 3 bars and a gas gain of 20000, normalized to the same area. The measurements with cosmic ray muons 
are compared to Garfield simulations\cite{garfield}.}
\label{fig_dtspectrum}
\end{figure}

A smaller outer tube diameter of 15~mm instead of 30~mm leads
to a reduction of the maximum drift time by a factor of 3.5 from about 700~ns to 200~ns 
(see Fig.~\ref{fig_dtspectrum}) when the operating voltage is reduced from 3080~V to 2730~V to keep the
gas gain the same. In addition, the background counting rate, dominated by the
conversion of the neutron and gamma radiation in the tube walls, is reduced by a factor of two
per unit tube length proportional to the tube circumference. Neglecting electronics shaping and dead time,
which should be minimized, both effects lead to a reduction of the drift-tube occupancy by about 
a factor of 7. The occupancy of 15~mm diameter drift tubes stays below $30\%$ up to the highest counting
rate of 1500 kHz expected in 2~m long tubes in the ATLAS inner forward chambers at S-LHC. A first verification of the
expected increase of the detection efficiency of 15~mm diameter compared to 30~mm diameter drift tubes
at high counting rates has been obtained up to rates of about 300 kHz per tube at a recent test with cosmic rays at the
CERN Gamma Irradiation Facility GIF~\cite{GIF}.

The space-charge distribution generated by the ion clouds drifting towards the tube wall changes the
electric field, influencing both the drift velocity and the gas gain. By lowering the effective potential experienced 
by the electrons drifting to the wire, high counting rates lead to decreasing gas gain. The resulting signal loss
grows with the inner tube radius $R_2$ proportional to $R_2^3\cdot \ln (R_2/R_1)$~\cite{riegler}, where $R_1=25~\mu$m
is the wire radius, and is therefore 10 times smaller in 15 mm compared to 30 mm diameter tubes.
Fluctuations of the space charge and of the electric field lead to variations of the drift velocity
causing a deterioration of the spatial resolution in non-linear drift gases like Ar:CO$_2$ (93:7)  
where the drift velocity depends on the electric field.
The latter effect increases strongly with drift distance above a value of about 7.5~mm while the gain drop effect
on the spatial resolution dominates for distances close to the sense wire. For drift radii below 7.5~mm
the space-to-drift time relationship is more linear leading to a reduced sensitivity to
environmental parameters such as gas composition and density, magnetic field and irradiation rate.

Since the spatial resolution of the drift tubes increases with the drift radius, the average single-tube
resolution deteriorates from $100~\mu$m for 30~mm diameter tubes~\cite{deile,sandra} to about $130~\mu$m
for 15~mm diameter tubes at low rates. The rate dependence of the 30~mm diameter drift-tube resolution
has been measured previously~\cite{deile,sandra} in a muon beam at the CERN GIF facility.
The resolution deteriorates linearly with the counting rate to about $120~\mu$m at 500~Hz/cm$^2$.
For 15~mm diameter tubes, the rate dependence of the resolution, dominated by the gain drop effect, is predicted
to be more than 10 times smaller causing a degradation of the resolution by only $20~\mu$m at the maximum
counting rate of 5000~Hz/cm$^2$ expected at S-LHC.

\section{Chamber design and fabrication}

The design of new drift-tube chambers with 15~mm diameter tubes follows as much as possible 
the current ATLAS MDT chamber design. In order to use the new chambers for an upgrade of the endcap region
of the ATLAS muon spectrometer, they have to fit into the same volume as the current chambers. This allows for
at least twice the number of drift-tube layers compared to the existing chambers with 30~mm diameter tubes and
a corresponding improvement of the track segment reconstruction efficiency and spatial resolution. The baseline
design of new inner forward chambers thus comprises two times eight tube layers compared to the two times four layers 
of the present inner forward MDT chambers. The spatial resolution of these chambers is 
expected to be better than $40~\mu$m up to a counting rate of 5~kHz/cm$^2$ with a sense wire positioning accuracy in the chambers 
of $50~\mu$m which is a less stringent requirement then for the current MDT chambers.

\begin{figure*}[!ht]
\centering
\includegraphics[width=\textwidth]{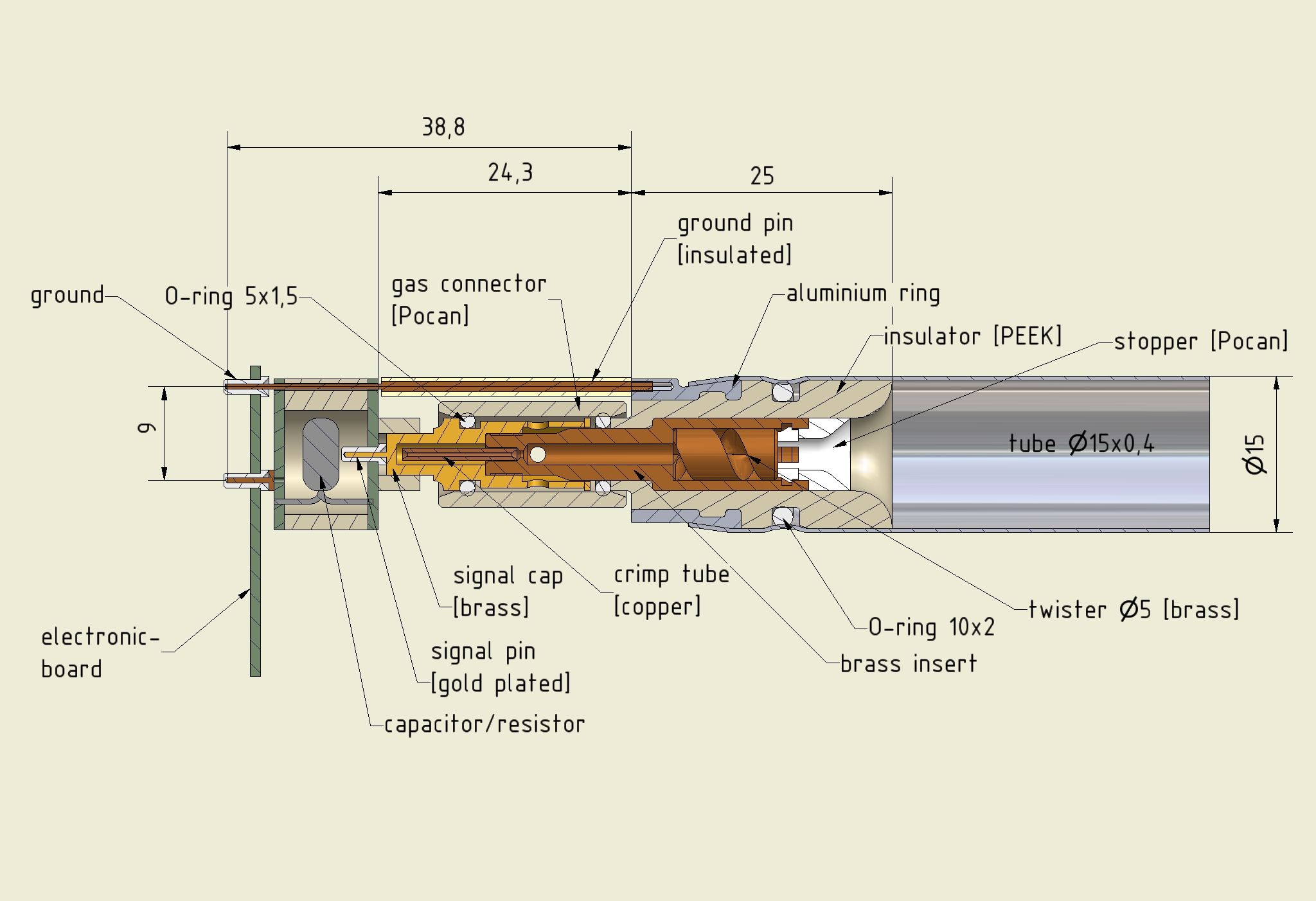} 
\caption{Endplug for 15~mm diameter drift tubes with interfaces to gas distribution system and electronics boards (see text).}
\label{fig_endplug} 
\end{figure*}

\begin{figure}[!h]
\centering
\includegraphics[width=0.5\textwidth]{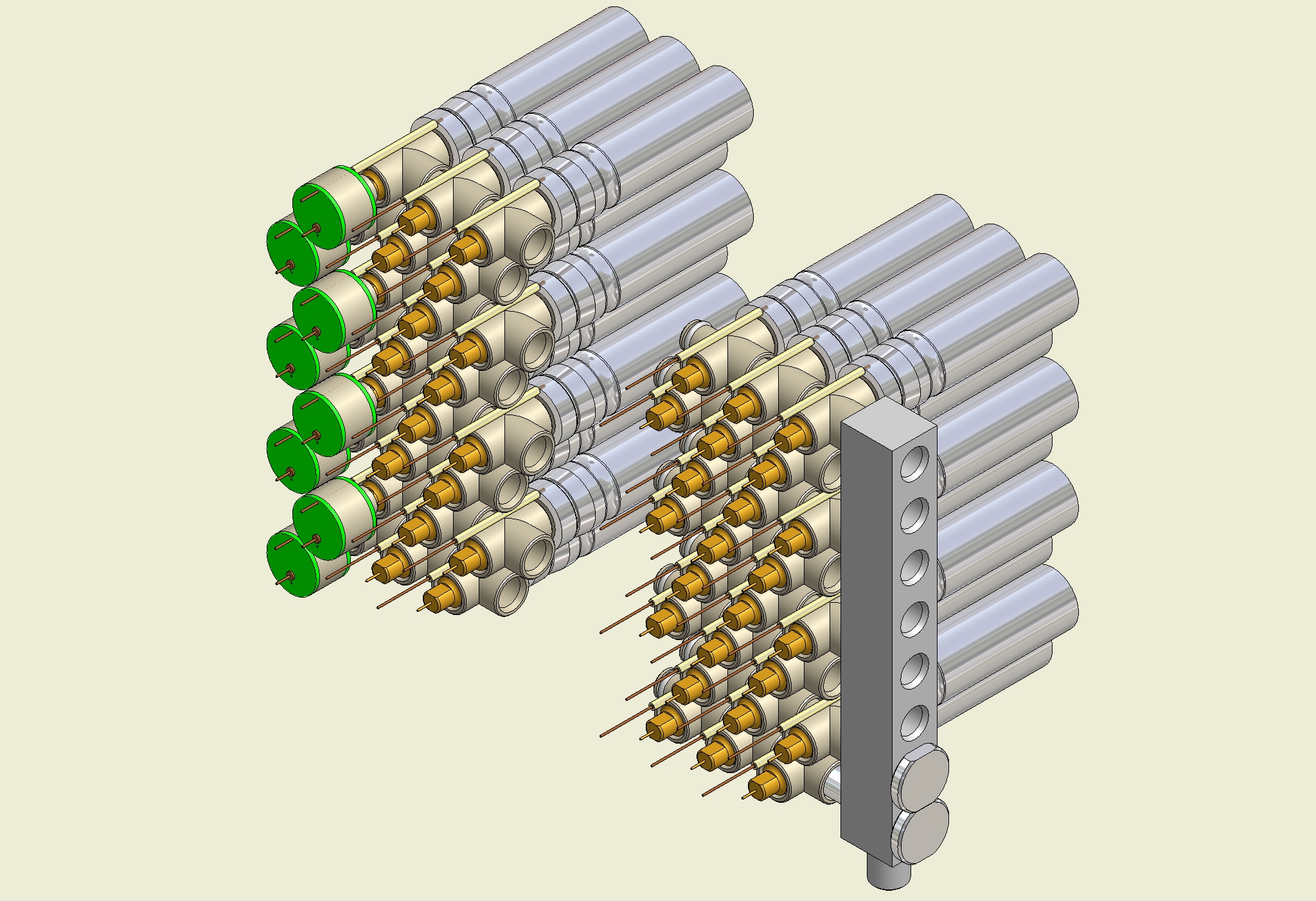}
\caption{Gas distribution system for a multilayer consisting of eight layers of 15~mm diameter drift tubes (see text).}
\label{fig_gasmanifold}
\end{figure}

The challenge for the new chamber design is the four times denser tube package with corresponding gas and electrical 
connections to the individual tubes.
Central to the chamber design is the development of the tube endplug (see Fig.~\ref{fig_endplug}) which insulates the sense wire from the tube wall,
centers the wire in the tubes with respect to an external reference surface on the endplug with an accuracy of about
$10~\mu$m and provides high-voltage-safe connections to the gas distribution manifolds (see Fig.~\ref{fig_gasmanifold}) and the readout and
high-voltage distribution boards. Gas leak rates have to stay below $10^{-8}$~bar$\cdot$l/s per tube 
in order to prevent contamination of the drift gas. The wire is fixed at both tube ends in copper crimping tubes inserted into the 
central brass inserts of the endplugs and connected to the signal and high-voltage distribution boards via brass signal
caps screwed onto the brass inserts and sealing the tubes with the gas manifolds at both ends with rubber O-rings.       
The brass insert holds the spiral-shaped wire locator (called twister) on the inside of the tube and transfers its position
in the plane perpendicular to the wire to a precisely machined reference surface on the outside of the tube which is used 
for the accurate relative positioning of the drift tubes in the chamber during chamber assembly. 
The endplugs are fabricated by injection molding and sealed in the tubes with O-rings by circular crimping of the tube walls.
Injection moulded adapter pieces, containing the high-voltage decoupling capacitor, connect the signal 
caps to the readout and high-voltage distribution boards at the two tube ends which become feasible in this design.

\begin{figure}[h]
\centering
\includegraphics[width=0.5\textwidth]{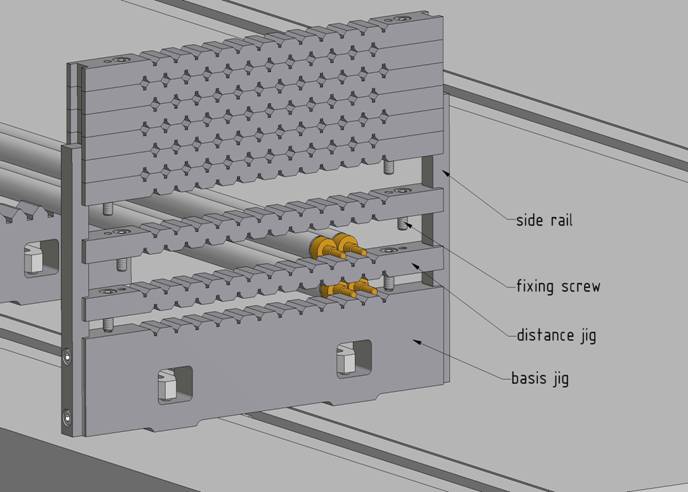}
\caption{Jigging for the assembly of a eight-layer package of 15~mm diameter aluminum drift tubes.}
\label{fig_jigging}
\end{figure}

The tubes are assembled to a chamber using the precision jigging tool shown in Fig.~\ref{fig_jigging}. The comb-shaped device
is machined with an accuracy of about $10~\mu$m and positions the reference surfaces of the tubes of a multilayer in the plane transverse to the wires. 
The eight tube layers of one multilayer are assembled and glued together in a single step
requiring only one working day. A module of 12x8 drift tubes has been assembled with this procedure. The external reference surfaces 
of the endplugs have been measured on a coordinate measuring machine after the glueing yielding the required accuracy of $50~\mu$m rms.
This has been achieved with 1~m long DIN standard aluminum tubes with tolerances of $\pm 0.1$~mm on diameter, roundness
and concentricity of inner and outer circumference and of $\pm 0.5$~mm on straightness.
After the assembly of the drift tubes, gas manifolds, signal caps and electronics boards are mounted.  
The whole chamber is finally enclosed in an aluminum faraday cage.

\begin{figure}
\centering
\includegraphics[width=0.5\textwidth]{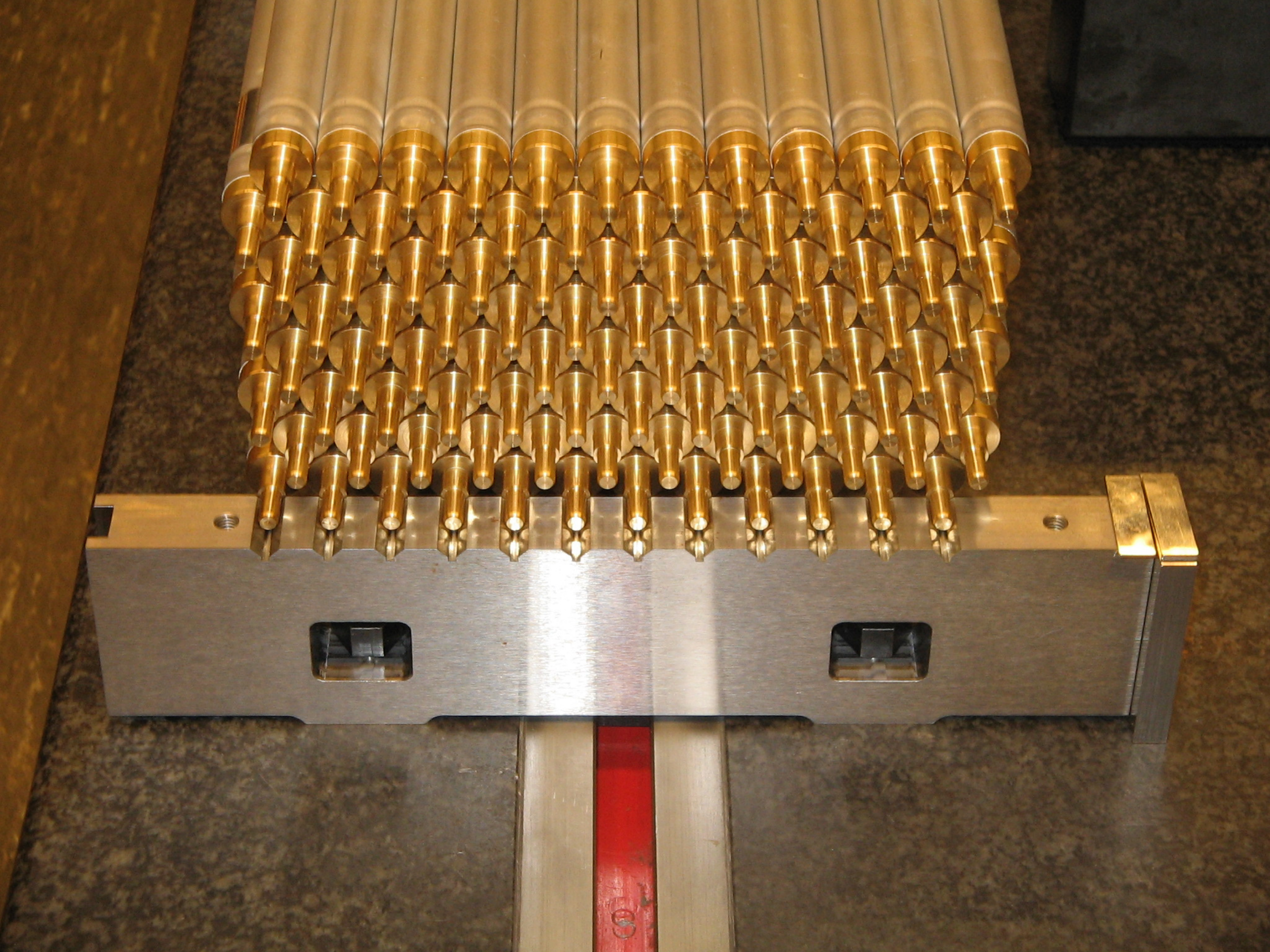}
\caption{Assembled 12 x 8 package of 15~mm diameter drift tubes.}
\end{figure}

\section{Conclusions}

Drift tube detectors provide robust and efficient tracking at high occupancies expected in the muon detectors
of the LHC experiments, in particular in the ATLAS detector. Drift tubes with 15~mm diameter
are sufficiently fast to cope with the counting rates expected in the forward regions of the ATLAS muon spectrometer
under the worst background conditions at LHC with ten times increased design luminosity (Super-LHC). 
The design of muon chambers with 15~mm diameter drift tubes is in an advanced stage. The construction of
a first prototype chamber is under preparation.


\begin{thebibliography}{1}

\bibitem{ATLAStp}
The ATLAS collaboration, \emph{ATLAS Technical Proposal}, CERN/LHCC
94-43, December 1994.
\bibitem{ATLASpaper}
The ATLAS collaboration, \emph{The ATLAS Experiment at the CERN LHC},
JINST 3 S080003 (2008).
\bibitem{MuonTDR}
The ATLAS collaboration, \emph{Technical Design Report for the ATLAS
Muon Spectrometer}, CERN/LHCC 97-22, May 1997.
\bibitem{baranov}
S. Baranov~{\it et al.}, \emph{Estimation of Radiation Background,
Impact on Detectors, Activation and Shielding Optimization in
ATLAS}, ATLAS internal note, ATL-GEN-2005-001 (2005).
\bibitem{aleksa}
M. Aleksa~{\it et al.}, \emph{Rate Effects in High-Resolution Drift
Chambers, Nucl}. Instr. and Meth. A 446 (2000) 435-443.
\bibitem{deile}
M. Deile~{\it et al.}, \emph{Performance of the ATLAS Precision Muon
Chambers under LHC Operating Conditions}, Nucl. Instr. and Meth.
A518 (2004) 65-68.
\bibitem{sandra}
S. Horvat~{\it et al.}, \emph{Operation of the ATLAS Precision Muon
Drift-Tube Chambers at High Background Rates and in Magnetic Fields},
IEEE trans. on Nucl. Science Instr. Vol. 53, 2 (2006) 562-566.
\bibitem{GIF}
J. Dubbert~{\it et al.}, \emph{Precision Drift Tube Chambers for the ATLAS Muon 
Spectrometer at Super-LHC}, proceedings of the 2008 IEEE Nuclear Science Symposium,
Dresden, Germany, 19-25 October 2008, Nuclear Science Symposium Conference Record 2008,
IEEE, 2008, MPI report, MPP-2008-191, November 2008.
\bibitem{riegler}
W. Riegler, \emph{High Accuracy Wire Chambers}, Nucl. Instr. and Meth. A494 (2002)
173-178.
\bibitem{garfield}
R. Veenhof, \emph{GARFIELD: Simulation of Gaseous Detectors, Version 8.01, CERN},
write-up: http://wwwinfo.cern.ch/writeup/garfield.



\end{thebibliography}
\end{document}